\documentstyle[aps,prl,twocolumn,rotating,epsfig]{revtex}
\begin{document}
\draft
\preprint{}

\title{Hysteresis and  Domain Formation in a Double-Layer Quantum Hall System at total Filling Factor 2}

\author{ J.G.S. Lok, A.K. Geim, B. Tieke, J.C. Maan}

\address{High Field Magnet Laboratory, Research Institute for Materials,
University of Nijmegen, NL-6525~ED Nijmegen, The~Netherlands\\}

\author{ S.T. Stoddart, R.J. Hyndman, B.L.~Gallagher, and M. Henini\\}

\address{Department of Physics, University of Nottingham, Nottingham NG7 2RD, England\\}

\date{\today}

\maketitle
\begin{abstract}
We report anomalous behavior in a double-layer two dimensional hole gas
(2DHG) at even integer filling factors which includes a hysteresis in the
longitudinal and Hall resistances and a very weak temperature dependence
of the resistance minima. All anomalies disappear and the conventional quantum
Hall effect behavior recovers when a thin metal film is
placed on the top of the 2DHG. The behavior is attributed to the presence
of the theoretically predicted magnetic ordering at even integer filling
factors which causes the formation of macroscopic spin-charge domains.
\end{abstract}


\narrowtext

Double-layer spin-polarized two dimensional systems have recently been under
intense scrutiny due to extraordinary magnetic properties they are expected to
exhibit~\cite{Zheng,DasSarma}. These properties arise from interlayer
Coulomb and exchange interactions. Depending in a subtle way on these
interactions as well as the Zeeman energy and the splitting between
subbands, both ferromagnetic and anti-ferromagnetic ordering between spins
in the layers has been predicted at filling factor~2.
\par
In this paper we report the first observation of hysteresis and domain
formation in such a system, a behavior most familiar for conventional
ferromagnetic materials. We have studied transport properties of a
$<$311$>$ grown double-layer two dimensional hole gas confined in two
100~\AA \ wide GaAs quantum wells separated by a 30 \AA \ thin AlAs barrier.
All results presented here are measured on a Hall bar along the $<$233$>$
direction, but similar results are obtained for Hall bars along the $<$011$>$
direction.
The 2DHG has a mobility of 11.1 (m$^2$/Vs) and a concentration of 1.8~10$^{15}$
(m$^{-2}$) equally distributed between the wells.
\par
The thin barrier in our structure gives rise to very strong interlayer
interactions and, at the same time, in contrast to the situation in a
double-layer {\em electron} gas, does not induce significant tunneling because
of the high effective mass of holes. This particularly implies a much smaller
splitting between the energies of the symmetric and anti-symmetric subbands
in the 2DHG.

\begin{figure}
\centering
\begin{turn}{270}
\epsfig{file=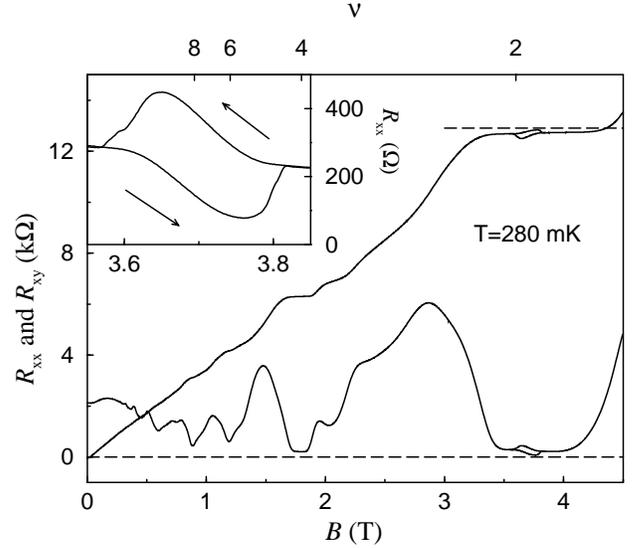, height=0.35\textheight,clip=,bbllx=69,bblly=104,bburx=535,bbury=628}
\end{turn}
\caption{Hall and longitudinal resistances of an ungated sample at a
temperature of 280 mK, showing poor quantization in Hall resistance and
considerable conductivity in the gap at filling factor~2. The inset shows
more pronounced the hysteresis in the longitudinal resistance near $\nu$=2.
Arrows indicate the direction of the magnetic field sweep.}
\label{figure1}
\end{figure}

\par
Figure~\ref{figure1} plots longitudinal and Hall resistances of our 2DHG
versus magnetic field at a temperature of 280 mK~\cite{Rhonda}.
The most striking feature in this figure is the pronounced hysteresis
in both R$_{xx}$(B) and R$_{xy}$(B) traces near filling factor~2. When sweeping
the magnetic up
R$_{xx}$ (R$_{xy}$) first decreases (increases), while on sweeping the magnetic
field down R$_{xx}$ (R$_{xy}$) first increases (decreases). The hysteresis
is observed at low temperatures (T $<$ 0.6~K) in both AC and DC measurements,
and its symmetry upon reversing the direction of the magnetic field is
symmetric for R$_{xx}$ and anti-symmetric for R$_{xy}$. The hysteresis 
is accompanied by a poorly quantized Hall resistance and a non-zero
longitudinal resistance, unexpected for such a high mobility 2DHG. Remarkably,
the hysteresis completely disappears and the quantization becomes exact when a
thin metal film (gate) is placed on top of the 2DHG.

\begin{figure}
\centering
\begin{turn}{270}
\epsfig{file=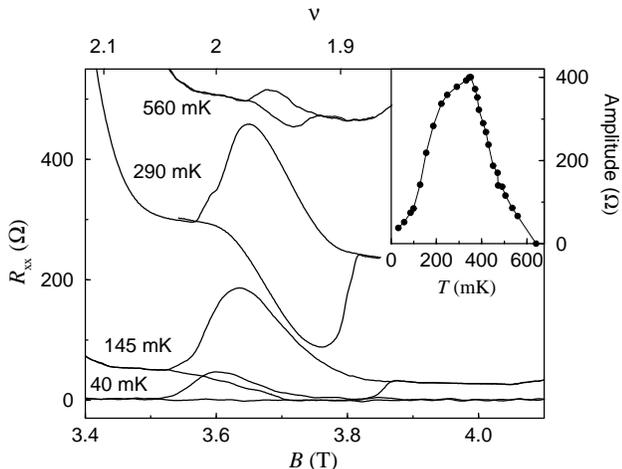, height=0.35\textheight,clip=,bbllx=69,bblly=90,bburx=535,bbury=702}
\end{turn}
\caption{Temperature dependence of the R$_{xx}$(B)-traces near $\nu$=2. Inset
plots the size of the hysteresis loop versus temperature.}
\label{figure2}
\end{figure}

\par
Figure~\ref{figure2} displays the magnetic field dependence of R$_{xx}$ near
$\nu$=2 for various temperatures.  At high temperatures there is no hysteresis
and R$_{xx}$ around $\nu$=2 decreases slightly with increasing magnetic field.
Below about 0.6~K, the hysteresis appears. Initially the size of the
hysteresis loop increases with decreasing temperature, until below 0.32~K the 
size decreases (see inset figure~\ref{figure2}). A part of the decrease below
0.20~K is due to R$_{xx}$ reaching zero on sweeping the magnetic field up.
\par
The hysteretic feature is not stable, i.e. it slowly relaxes towards the
middle of the hysteresis loop when the magnetic field is kept constant.
Initially the time evolution of the signal is steeper than exponential,
while for longer times it is much slower than exponential. At 0.50~K
the lifetime of the hysteresis (defined as the time it takes to reach a
resistance within 10 \% of the value at equilibrium) is a few minutes,
while at 40~mK the equilibrium cannot be reached within 3 hours (our longest
experiment). The width and the size of the hysteresis loop depend
slightly on sweep rate (the size
decreases by 60 \% on reducing the sweep rate by two orders of magnitude
to 0.001 T/min), but this is solely due to the relaxation occurring
during such slow sweeps and not due to the change of dB/dt.
\par
\begin{figure}
\centering
\epsfig{file=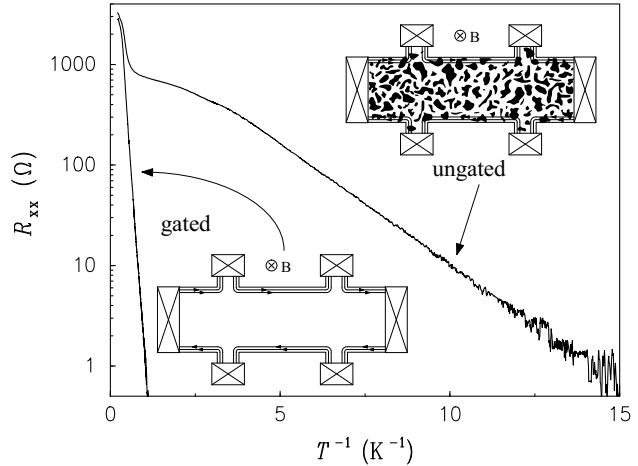, height=0.27\textheight,clip=,bbllx=29,bblly=21,bburx=300,bbury=224}
\caption{Longitudinal resistance versus inverse temperature at filling factor~2
for the gated and the ungated sample. Activation energy for the gated
sample is 10.5~K and for the ungated sample 0.56~K (below 0.30~K). While 
in gated samples at $\nu$=2 edge channels propagate along the boundary of the 
sample, in ungated samples these edge channels are backscattered by the
macroscopic density fluctuations (schematically indicated by the shaded
regions).
}
\label{figure3}
\end{figure}

Figure~\ref{figure3} compares temperature dependences of the longitudinal
resistance of a gated and an ungated sample at total filling factor~2. For the
gated structure, R$_{xx}$ is thermally activated with an activation energy of
10.5~K in good agreement with the behavior of conventional quantum Hall effect
(QHE) structures. Initially, the ungated sample follows the same activation
curve, until below $\sim$2~K it starts to flatten. At even lower temperatures
(T $<$ 0.3~K) it is activated again, but with a much lower activation energy
of only 0.56~K. During a temperature sweep, the measured R$_{xx}$ corresponds
to the middle of the hysteresis loop, i.e. it approximately
falls on the line connecting the points at which the loop opens.

\par
\begin{figure}
\centering
\epsfig{file=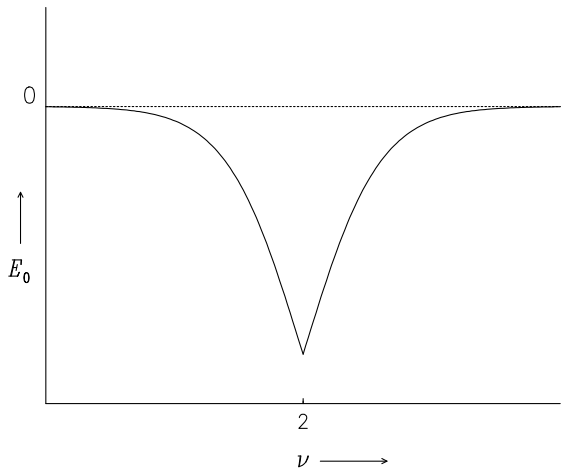, height=0.27\textheight,clip=,bbllx=474,bblly=21,bburx=715,bbury=222}
\caption{Phenomenological model to explain domain formation in our ungated
structures.}
\label{figure4}
\end{figure}

Both the non-zero longitudinal resistance as well as the poorly quantized
Hall resistance indicate the presence of significant fluctuations in the
ungated 2DHG which give rise to backscattering and bulk
conductivity~\cite{Haug}. Their absence in the presence of a nearby metal plate
is presumably due to screening and implies the electrical nature of the
fluctuations. To explain such fluctuations, we employ
the theoretical prediction that strongly coupled double-layer quantum Hall
systems at $\nu$=2 exhibit transitions to an (anti-)ferromagnetic ground
state and that away from this filling factor the ground state is
destroyed~\cite{Zheng,DasSarma}(see figure~\ref{figure4}). Then, at
a magnetic
field slightly away from $\nu$=2, the energy gain due to the magnetic transition
drives the system towards segregation into two macroscopic phases: the
energetically favorable phase with $\nu$=2 and the rest of the sample moved
further away from this filling factor (see figure~\ref{figure3}). The phases
differ in their carrier concentrations in such a way as to keep the average
value constant and the size of the domains depends on the energy gained in
the magnetic transition and the energy lost due to stray electric fields.
Since in gated samples the metal plate is 170 nm away from the top quantum
well, the minimum size of the domains must be at least of this order of
magnitude. At the lowest temperatures the longitudinal resistance in ungated
samples is thermally activated again, which suggests the presence of a well
defined gap separating the (anti)-ferromagnetic ground state responsible for
the domain formation, from the Fermi energy.

\par
The above model also accounts for the behavior of the hysteresis which, as the
experiment shows, is a non-equilibrium state of the system. The key-point
is that the initial states just below and just above $\nu$=2 are not
identical. When approaching $\nu$=2 from the low magnetic field side, the
sample initially breaks up into domains with a {\em lower} density and the rest
of the sample with a {\em higher} density. When approaching from the high field
side, this is the other way around. In the former case, the topmost edge
channel inside the domains can freely propagate through the rest of the sample,
while in the latter case, this topmost edge channel is strongly reflected
by the lower density inclusions as it is known to be the case in experiments
with selectively gated QHE samples~\cite{Haug}.
\par
The slow relaxation of the hysteresis is likely to be due to changes in
size and/or position of the domains so as to reach equilibrium.
At the lowest temperatures the lifetime of the hysteresis becomes extremely
long, either because the energy barrier for breaking up domains becomes
larger than the thermal energy (as in ferromagnets) or because random
potential fluctuations in the sample are larger than the thermal energy and
pin the domains.
\par
Finally, we want to mention that similar behavior has been observed at
$\nu$=4, although hysteresis appears at a much lower temperature (80~mK).
At $\nu$=4 the activation energy in gated samples is 3.7~K, while that in
ungated samples is only 0.27~K.
\par
In conclusion we presented experimental evidence for the formation of
macroscopic domains in a strongly coupled 2DHG at total filling factor~2.
The pronounced hysteresis observed in the longitudinal and Hall resistances
near filling factor~2, is a non-equilibrium state, but one with an extremely
long lifetime at low temperatures. A phenomenological model
accounts for both the formation of these domains as well as the observed
hysteresis.
\par
\hyphenation{Stich-ting}
This work is part of a research program of the Stichting voor Fundamenteel
Onderzoek der Materie (FOM) financially supported by NWO (The Netherlands).

\end{document}